\documentclass[aps,prl,showpacs,reprint,groupedaddress]{revtex4-1}

\usepackage{mathtools}
\usepackage{bbold}
\usepackage{eqnarray,amsmath,amscd,amssymb,graphicx}
\usepackage[lofdepth,lotdepth,caption=false]{subfig}
\usepackage{textcomp}
\usepackage{natbib}

\newcommand{\degp}{\ensuremath{{\cal P}}}
\newcommand{\g}[1]{\ensuremath{\mathbf{#1}}}
\newcommand{\mat}[1]{\ensuremath{\uline{\mathrm{#1}}}}
\newcommand{\rom}[1]{\ensuremath{{\mathrm{#1}}}}
\newcommand{\Id}{\ensuremath{\uline{\mathrm{Id}}}}
\newcommand{\tr}{\ensuremath{\mathrm{tr}}}
\newtheorem{prop}{Property}
\newcommand{\mink}[1]{\ensuremath{||#1||_{1,3}}}
\usepackage{ulem}
\usepackage{cancel}
\usepackage{color}
\definecolor{vert}{rgb}{0.0,0.4,0.0}
\definecolor{rouge}{rgb}{0.6,0.0,0.0}

\renewcommand{\emph}[1]{\textit{#1}}

\newcommand*{\diag}{\mathrm{diag}}

\begin{document}

\title{Differential description and irreversibility  of
  depolarizing light-matter interactions}

\author{Julien Fade}
\email{julien.fade@univ-rennes1.fr\\}
\affiliation{Institut de Physique de Rennes, CNRS, Universit\'e de Rennes 1, Campus de Beaulieu, 35\,042 Rennes, France}
\author{No\'e Ortega-Quijano}
\affiliation{Institut de Physique de Rennes, CNRS, Universit\'e de Rennes 1, Campus de Beaulieu, 35\,042 Rennes, France}
\date{\today}

\begin{abstract}

  The widely-used Jones and Mueller differential polarization calculi
  allow non-depolarizing deterministic polarization interactions,
  known to be elements of the $SO^+(1,3)$ Lorentz group, to be
  described in an efficient way. In this Letter, a stochastic
  differential Jones formalism is shown to provide a clear physical
  insight on light depolarization, which arises from the interaction
  of polarized light with a random medium showing fluctuating
  anisotropic properties. Based on this formalism, several
  \emph{intrinsic} depolarization metrics naturally arise to
  efficiently characterize light depolarization in a medium, and an
  irreversibility property of depolarizing transformations is finally
  established.

\end{abstract}


% 02.20.Sv 	Lie algebras of Lie groups
% 78.20.Bh 	Theory, models, and numerical simulation
% 42.25.Bs 	Wave propagation, transmission and absorption
% 42.25.Ja 	Polarization
% 89.70.Cf      Entropy in information theory
%\ocis{}

\maketitle
In the field of polarimetry, Jones and Stokes/Mueller formalisms have
always appeared as dual and often exclusive approaches, whose specific
characteristics have been exploited for diverse applications. On the
one hand, the description of field coherence in the Jones calculus,
which relates the input and output 2-dimensional complex electric
field through $\g{E}_{out}=\mat{J}\g{E}_{in}$, justifies its use in
ellipsometry \cite{azz87,web10}, optical design
\cite{chi89,kam81,col93,kli12}, spectroscopy \cite{kli12}, astronomy
\cite{cou94} or radar (PolSar) \cite{lee09}. On the other hand,
Mueller calculus is widely used in biophotonics \cite{pie11,ala15},
material characterization \cite{rog11, mag14} or teledetection
\cite{tyo06}, as it is based on optical field observables (intensity
measurements), relating the input and output 4-dimensional real Stokes
vector through $\g{s}_{out}=\mat{M}\g{s}_{in}$. As a consequence,
these approaches fundamentally differ in their capacity to
characterize depolarizing light-matter interactions (i.e.,
non-deterministic polarization transformations yielding a partial
randomization of the input electric field). As Jones already pointed
out in one of his seminal papers \cite{jon47}, Jones matrices are
unable to directly describe depolarizing media, which can however be
grasped in the Mueller formalism via depolarizing Mueller
matrices. This discrepancy between both standpoints takes part in the
debate, still topical in the scientific community, about the physical
origin of light depolarization in media
\cite{bic92,bro98,aie05,xu05,set08,mac11,sor11,pou12,dev13c}.  In this
Letter, we show that the differential polarization formalism, which
naturally arises from group theory, provides new physical insight on
depolarizing light-matter interactions. This approach allows us to
define \emph{intrinsic} depolarization metrics, and to demonstrate an
irreversibility property for depolarizing transformations, as a
counterpart to the well-known invariance property verified by
deterministic interactions.

\begin{figure}[ht]
\begin{center}
\includegraphics[width=7.5cm]{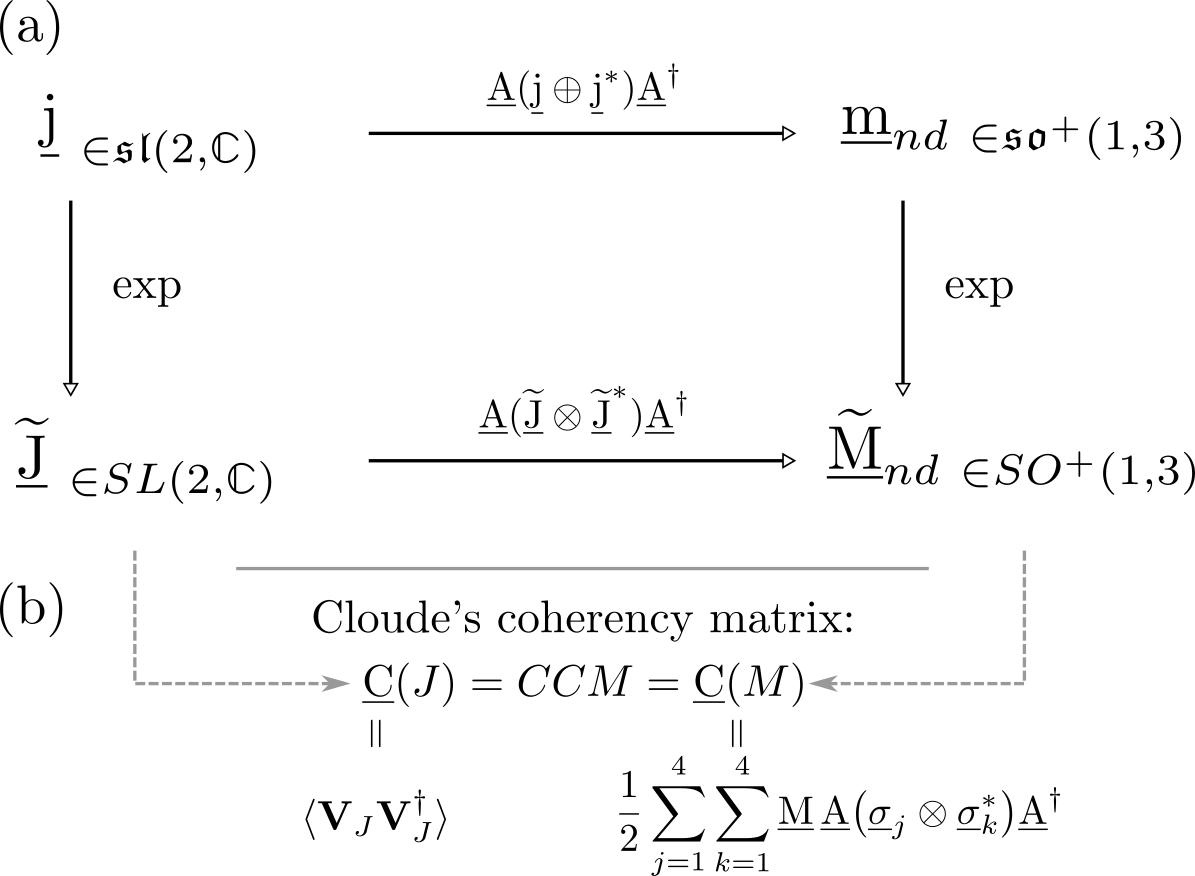}
\end{center}
\caption{(a) Commutative diagram linking differential and macroscopic
  Jones and (non-depolarizing) Mueller matrices, with
  $A=\bigl[\mathrm{vec}(\mat{\sigma}_0) \ \mathrm{vec}(\mat{\sigma}_1)
  \ \mathrm{vec}(\mat{\sigma}_2) \ \mathrm{vec}(\mat{\sigma}_3)
  \bigr]^T/\sqrt{2}$ and $A^{-1}=A^\dagger$. (b) Definition of
  Cloude's coherency matrix (CCM). Superscripts $*$, $T$ and
    $\dagger$ respectively denote complex conjugation, standard and
    Hermitian matrix transposition.\label{diag}}
\end{figure}

In the specific situation of a deterministic polarization
transformation, there is a clear one-to-one relationship (recalled in
Fig.~\ref{diag}) between a $2\times 2$ complex Jones matrix $\mat{J}$
and the corresponding $4\times 4$ real-valued non-depolarizing
Mueller-Jones matrix $\mat{M}_{nd}$ \cite{sim82,
  kim87}. Interestingly, when one considers normalized
unit-determinant matrices, both descriptions appear to be isomorphic
representations of the same 6-dimensional group, namely the proper
orthochronous Lorentz group $SO^+(1,3)$ for  unit-determinant
Mueller matrices $\widetilde{\mat{M}}_{nd}$ and the special linear
group $SL(2, \mathbb{C})$ for  unit-determinant Jones matrices
$\widetilde{\mat{J}}$ \cite{clo86, han97}. As a result, there is a
well-known analogy between deterministic polarization transformations
and special relativity
\cite{pel92,han97,mor03b,sim10,bas12,tud15,fra15}, as non-depolarizing
interactions correspond to Lorentz transformations, and must therefore
preserve the Minkowski metric (defined as
$\mink{\g{s}}^2=\g{s}^T G \g{s}$, with $G=\diag[1,-1,-1,-1]$) of the
input Stokes vector $\g{s}_{in}$ \cite{sim10}. Interestingly, this
invariance property can be related to the preservation of the Shannon
entropy of the field, under the assumption of complex Gaussian
circular fluctuations \cite{ref04b}, as
$H(\g{s})=-\int P_\g{E} (\g{E}) \ln P_\g{E}(\g{E})= \ln \pi^2 e^2
\mink{\g{s}}^2/4$. As pointed out in \cite{fra15}, such an invariance
property is neither verified by the intensity of the light , $I=s_0$,
nor by its degree of polarization,
$\degp = \sqrt{s_1^2+s_2^2+s_3^2}/s_0$.

Historically, Jones \cite{jon48} and Azzam \cite{azz78} respectively
introduced the so-called \emph{differential} Jones and Mueller
calculi, with a corresponding differential Jones matrix (dJm)
$\mat{j}$ and a differential Mueller matrix (dMm) $\mat{m}_{nd}$. Both
approaches describe the local evolution of a transversally polarized
wave along direction ${\hat{\g{n}}}$ through the respective
differential equations
% \begin{equation}
$ d\g{E}/d{\hat{\g{n}}}=\mat{j}\, \g{E}$, and
$ d\g{s}/d{\hat{\g{n}}}=\mat{m}\, \g{s}$.
 %\end{equation}
According to group theory, these differential descriptions lead to a
representation of deterministic polarization transformations (either
in group $SL(2,\mathbb{C})$ or in $SO^+(1,3)$), by their counterpart
in the corresponding Lie algebra, i.e., $\mathfrak{sl}(2,\mathbb{C})$
for $\mat{j}$, or $\mathfrak{so}^+(1,3)$ for $\mat{m}_{nd}$, which
verifies Minkowski G-antisymmetry with
$\mat{m}_{nd} +\mat{G}\,\mat{m}_{nd}^T\,\mat{G}=0$ \cite{lou01}. There
is a clear equivalence between these four representations, which are
linked through the commutative diagram represented in Fig.~\ref{diag},
the macroscopic and differential matrices being related by the
exponential map ($\mat{J}=\exp( \mat{j}\Delta z)$ and
$\mat{M}_{nd}=\exp( \mat{m}_{nd} \Delta z)$ when 
propagation over $\Delta z$ through a homogeneous medium is assumed).

As Lie algebras can be viewed as the tangent spaces to the
corresponding Lie groups at the identity element \cite{gal11,lou01},
the differential Jones or Mueller formalisms allow polarization
properties to be described in a linearized geometry. One of the
powerful consequences of such a linearization lies in the simple
linear parameterizations of the differential matrices in terms of
anisotropic optical properties. Indeed, the dJm $\mat{j}$ and dMm
$\mat{m}_{nd}$, that both characterize the polarimetric properties of
an infinitesimal plane-parallel slab of a deterministic linear optical
medium, read
\begin{equation}
\mat{j}=\frac{1}{2}
 \begin{bmatrix}\label{defj}
  \kappa_i+ \kappa_q - i (2 \eta_i + \eta_q) &\kappa_u - \eta_v - i ( \eta_u+ \kappa_v )\\
\kappa_u + \eta_v - i (\eta_u -\kappa_v  )&  \kappa_i- \kappa_q - i (2 \eta_i - \eta_q)
\end{bmatrix}, 
\end{equation}
\begin{equation}\label{defmnd}
\text{and,} \quad \mat{m}_{nd}=
\begin{bmatrix}
\kappa_i& \kappa_q & \kappa_u & \kappa_v \\
\kappa_q&\kappa_i  &-\eta_v &\eta_u\\
\kappa_u&\eta_v&\kappa_i  &-\eta_q\\
\kappa_v&-\eta_u&\eta_q&\kappa_i 
\end{bmatrix}.
\end{equation} 
Notations $\kappa$ and $\eta$ respectively refer to the real and
imaginary part of the propagation constant, which reads
$p=\kappa - i \eta$ when the considered monochromatic plane wave is
written as $\g{E}=\g{E}_0 \exp[-(i\omega t + p z)]$
\cite{jon48,mul69}. In the above matrices describing anisotropic
optical media, the isotropic extinction coefficient is denoted
$\kappa_i$, whereas $\eta_i$ stands for the isotropic (absolute)
optical phase, whose information is lost in the Mueller description
\cite{jon47}. As for the other terms, the subscripts $q$, $u$, and $v$
refer to linear $x$-$y$, linear $\pm 45^\circ$ and circular
\emph{left/right} anisotropy, through
$x_{q,u,v}=x_{x,45^\circ,rcp} - x_{y,-45^\circ,lcp} $, for $x=\kappa$
or $\eta$.

One can now legitimately question the physical understanding of
depolarizing transformations, which entail randomized anisotropic
effects on the incident optical wave.  Recently, it has been proposed
to extend the differential Mueller formalism to the more intricate
case of depolarizing transformations, by the introduction of
depolarizing dMm's. This approach has permitted a number of
interesting results to be obtained on depolarizing transformations
\cite{ort12,kum12,ort14,ell14,vil14,oss14b,ort15,oss15,aga15}.  We
propose below to theoretically justify the recent developments on
depolarizing dMms, by using an alternative description involving
stochastic differential Jones matrices. For that purpose, let us
consider a stochastic differential Jones matrix
$\mat{j}=\mat{j}_0+ \Delta \mat{j}$, modeling a random depolarizing
local transformation of the field, where
$\mat{j}_0=\langle \mat{j}\rangle $ is the deterministic average
polarization transformation, whose form has been recalled in
Eq.(\ref{defj}), and where the fluctuations matrix verifies
$\langle \Delta \mat{j} \rangle=0$. Assuming infinitesimal propagation
over $\Delta z$, the Jones matrix for such a transformation can be
written
$ \mat{J} =\exp(\mat{j}\Delta z) \simeq \Id +\mat{j} \Delta z $ at
first order in $\Delta z$. From Eq.~(\ref{defj}), this relation can be
conveniently rewritten in a vector form in the Pauli matrices basis
$\{\g{\sigma}_i\}_{i=\{0,3\}}$ as
 % \begin{equation}\label{VJ}
 %   \g{V}_{\rom{J}} \simeq \frac{\Delta z}{2}\begin{bmatrix}\frac{2}{\Delta z}+ \kappa_i - 2  i \eta_i \\\kappa_q - i \eta_q \\  \kappa_u - i \eta_u\\ \kappa_v - i \eta_v\end{bmatrix}=\begin{bmatrix}1+ \frac{\Delta z}{2}p^{(0)}\\ \frac{\Delta z}{2}\g{p}\end{bmatrix},
 % \end{equation}
$\g{V}_{\rom{J}} \simeq \bigl[ 1+ \frac{\Delta z}{2}p^{(0)} \ \ \frac{\Delta
  z}{2}\g{p}^\dagger \bigr]^T $,
with $p^{(0)}=p^{(0)}_0+\Delta p^{(0)}$, $\g{p}=\g{p}_0+\Delta \g{p}$, and where 
 \begin{equation}\label{VJ}
p^{(0)}_0=\kappa_i - 2\eta_i,\text{ and } \g{p}_0=\begin{bmatrix}\kappa_q - i \eta_q \\  \kappa_u - i \eta_u\\ \kappa_v - i \eta_v\end{bmatrix}.
 \end{equation}

 From $\g{V}_{\rom{J}}$, one can derive the Cloude's coherency matrix
 (CCM) of the polarimetric transformation, which provides relevant
 criteria to assess the physical realizability of macroscopic Mueller
 matrices \cite{clo86,clo90}, by calculating
 $\mat{C}(\rom{J})=\langle \g{V}_{\rom{J}}
 \g{V}_{\rom{J}}^\dagger\rangle$, which can be decomposed into a sum
 of two terms $ \mat{C}(\rom{J}) = \mat{C}_{nd} + \mat{C}_d$, where
 $\mat{C}_{nd} = \g{V}_{\rom{J}_0} \g{V}_{\rom{J}_0}^\dagger$ with
$\g{V}_{\rom{J}_0} \simeq \bigl[ 1+ \frac{\Delta z}{2}p_0^{(0)} \ \ \frac{\Delta
  z}{2}\g{p}_0^\dagger \bigr]^T $, and where
\begin{equation}\label{C_Jtilde}
   \mat{C}_d=\frac{(\Delta z)^2 }{4}\begin{bmatrix}c_0 &
    \g{c}^\dagger\\\g{c} & \mat{{\cal C}} \end{bmatrix},
\end{equation}
with $c_0=\langle |\Delta p^{(0)}|^2\rangle$,
$\g{c}=\langle \Delta p^{(0)*} \Delta \g{p}\rangle $ and the
$3\times 3$ matrix
$\mat{{\cal C}}=\langle \Delta \g{p}\Delta \g{p}^\dagger \rangle$.
From such a decomposition, it is obviously seen that a deterministic
transformation, with $\Delta p^{(0)}=0$ and $\Delta \g{p}=\g{0}$,
results in a CCM of rank one ($\mat{C}_{nd}$ being the matrix of a
projector) which is Cloude's condition for a non-depolarizing
transformation \cite{clo86,clo90}. Conversely, as soon as
$\mat{{\cal C}}$ is a non-null matrix, the rank of $\mat{C}(\rom{J})$
is greater than one, hence the corresponding transformation is
depolarizing according to Cloude's criterion \cite{clo86,clo90}. As a
result, the depolarizing nature of a transformation appears to be
completely comprehended by the $3\times 3$ positive semi-definite
Hermitian submatrix $\mat{{\cal C}}$, i.e., by 9 independent real
parameters. As will be seen below, this matrix allows one to define
interesting \emph{intrinsic} depolarization metrics of the medium.

% In order to link the above results with the dMm description, it must
% be noted here that these 9 depolarization parameters correspond to a
% subset of the 10 complex variance/covariance terms (16 real terms) of
% $\mat{C}_d$ in a 4-dimensional space, obtained when absolute
% phase/intensity fluctuations are neglected, as in the Mueller
% formalism. As a result in the remainder of this Letter, we shall
% describe a polarimetric transformation up to a constant phase factor,
% i.e. we assume $p^{(0)}=\kappa_i$, and we neglect the absolute phase
% and intensity fluctuations by setting $\Delta p^{(0)}=0$. Under these
% conditions, $c_0=0$ and $\g{c}=\g{0}$ in Eq.(\ref{}).

%% thus reads
%% \begin{equation*}
%%   \mat{C}(\rom{J}) =\begin{bmatrix}1 + \frac{\Delta z}{2}\kappa_i \\ \frac{\Delta z}{2}\g{p}_0\end{bmatrix}.\begin{bmatrix}1+\frac{\Delta z}{2}\kappa_i& \frac{\Delta z}{2}\g{p}_0^\dagger\end{bmatrix}+\Bigl(\frac{\Delta z}{2}\Bigr)^2\begin{bmatrix}0 &
%%     \g{0}^\dagger\\ \g{0} & \mat{{\cal C}} \end{bmatrix}.
%% \end{equation*}

It is now quite straightforward to identify these terms with the CCM
of, respectively, the non-depolarizing and the depolarizing dMm's that
have been introduced in earlier works \cite{ort11b,
  ort11c,oss11}. Indeed, in the dMm formalism, the Mueller matrix for
the considered local transformation reads
$\mat{M} = \exp( \mat{m} \Delta z) \simeq \Id +\mat{m} \Delta z $ at
first order in $\Delta z$. As suggested in \cite{oss11}, the dMm
$\mat{m}$ can be decomposed into a G-antisymmetric part, namely
$\mat{m}_{nd}$ given in Eq.~(\ref{defmnd}), and a G-symmetric part
$\mat{m}'_{d}$ which corresponds to the depolarizing contribution.
With such a decomposition, it can be checked that the $3\times 3$
lower-right submatrix of the CMM of $\mat{M}$ is only due to the
G-symmetric (depolarizing) part of $\mat{m}$. As a result, since
Eq.~(\ref{C_Jtilde}) indicates that this submatrix must have a
quadratic behavior in $\Delta z$, the parameterization of the dMm must
be written
$\mat{m} = \mat{m}_{nd} + \mat{m}'_{d}=\mat{m}_{nd} + \mat{m}_{d}
\Delta z $, with the 9-parameters G-symmetric part reading
\cite{ort11b, ort11c, oss11}
\begin{equation}
\mat{m}_{d}=
\begin{bmatrix}
0 & d_{\kappa_q} &  d_{\kappa_u} &  d_{\kappa_v} \\
- d_{\kappa_q}&- d_{\mu_q}  &  d_{\eta_v} & d_{\eta_u}\\
-d_{\kappa_u}& d_{\eta_v}& -d_{\mu_u}  & d_{\eta_q}\\
- d_{\kappa_v}& d_{\eta_u}& d_{\eta_q}&- d_{\mu_v}
\end{bmatrix}.
\end{equation}
The proposed decomposition of $\mat{m}$ has an important physical
meaning: the depolarization properties of a sample must pile up
quadratically with $\Delta z$, whereas deterministic anisotropy
parameters classically evolve linearly with the propagation
distance. Such a decomposition has been recently proposed in
\cite{oss14}, but without a clear physical justification that
 is brought by the approach presented in this Letter. This
interesting property of depolarization in samples has been recently
verified experimentally on controlled test samples \cite{aga15}, and
it may have crucial implications in the analysis of
depolarizing media in experimental polarimetry \cite{web10,mar03}.

With such a parameterization, the CMM of
$\mat{M}\simeq \Id+ \mat{m}\Delta z$ is obtained using
the relationship recalled in Fig.~\ref{diag}, yielding
$ \mat{C}(\rom{M}) =\begin{bmatrix}1+ \kappa_i \Delta z & \frac{\Delta
    z}{2} \g{p}_0^\dagger \\ \frac{\Delta z}{2} \g{p}_0 &
  \bigl(\frac{\Delta z}{2}\bigr)^2 \mat{\Sigma} \end{bmatrix}$, with
\begin{equation*}
  \mat{\Sigma}=2 \begin{bmatrix}\frac{-d_{\mu_q} +d_{\mu_u}+d_{\mu_v} }{2}&d_{\eta_v} +i d_{\kappa_v} & d_{\eta_u} -id_{\kappa_u}\\
    d_{\eta_v} -i d_{\kappa_v}&\frac{d_{\mu_q} -d_{\mu_u}+d_{\mu_v} }{2}&d_{\eta_q} +id_{\kappa_q}\\
    d_{\eta_u} +id_{\kappa_u}&d_{\eta_q} -id_{\kappa_q} &\frac{d_{\mu_q} +d_{\mu_u}-d_{\mu_v} }{2} \end{bmatrix}.
\end{equation*}
It can be observed that $\mat{C}(\rom{M})$ corresponds to an
approximation of the CMM obtained above from the Jones formalism,
where each element has been truncated to the first non-null term of
its Taylor expansion in $\Delta z$.

It is now interesting to identify the lower-right $3 \times 3$
submatrix $\mat{\Sigma}$ with its previous expression
$\mat{\cal C}=\langle\Delta \g{p}\Delta \g{p}^\dagger\rangle$ found in
Eq.~(\ref{C_Jtilde}), yielding the following set of equations:
\begin{equation*}\begin{split}
2d_{\mu_{q,u,v}} &= \bigl\langle   [(\Delta \eta)^2 + (\Delta \kappa)^2]_{u,v,q} +  [(\Delta \eta)^2 + (\Delta \kappa)^2]_{v,q,u}  \bigl\rangle,\\
2d_{\eta_{q,u,v}}&=\bigl\langle \Delta \kappa_{u,v,q}\Delta \kappa_{v,q,u}+\Delta \eta_{u,v,q}\Delta \eta_{v,q,u} \bigl\rangle,\\
 2d_{\kappa_{q,u,v}}&=\bigl\langle \Delta \kappa_{u,v,q}\Delta \eta_{v,q,u}-\Delta \eta_{u,v,q}\Delta \kappa_{v,q,u}\bigl\rangle.
\end{split}\end{equation*}

This clearly shows that the nine depolarization parameters
$d_{\mu_{q,u,v}} $, $d_{\eta_{q,u,v}} $ and $d_{\kappa_{q,u,v}}$ of
the dMm $\mat{m}_d$ are physically related to the second-order
statistical properties of the anisotropy parameters of the
sample. Such an observation was recently reported for the first
  time in \cite{dev13c} through somewhat intricate calculus using
  stochastic dMm's. More interestingly, the fact that these nine
parameters are related to variance/covariance terms through the above
specific relationships is the fundamental origin of the necessary
conditions that must be fulfilled by the elements of $\mat{m}_d$ so
that it is physically admissible \cite{oss14,oss14b}. These necessary
conditions imply that an admissible depolarizing dMm $\mat{m}_d$ must
belong (up to double cosetting by Lorentz transformations) to one of
the two canonical forms derived in \cite{dev13,dev13b}:
\begin{subequations}\label{canon}
\begin{equation}\begin{split}
\mat{m}_d^{(\textsc{\romannumeral 1})}=\mathrm{diag} &\bigl[d_1+d_2+d_3;\ d_1-d_2-d_3;\\
&-d_1+d_2-d_3;\ -d_1-d_2+d_3\bigr],
\end{split}\end{equation}
\begin{equation}
\mat{m}_d^{(\textsc{\romannumeral 2})}=
\begin{bmatrix}
d_1+d_2&d_2&0&0\\
-d_2&d_1-d_2&0&0\\
0&0&-d_1&0\\
0&0&0&-d_1
\end{bmatrix},
\end{equation}
\end{subequations}
with the following conditions on the canonical depolarization
parameters: $d_i \geq 0$, for $i\in \{1,3\}$.

These previous results evidence the fact that the depolarization
properties of a medium at an infinitesimal level are intrinsically
described by the matrix $\mat{\Sigma}$ (or equivalently
$\mat{\cal C}$), which contains the 9 depolarization parameters
described above. However, standard depolarization metrics are defined
either on the macroscopic Mueller matrix of the medium (e.g.,
depolarization index \cite{gil86}), or on its CCM (e.g., Cloude
entropy \cite{clo86,clo90,clo95}). Though often useful, such
depolarization metric definitions can nevertheless be unsatisfactory
in some situations. Indeed, two samples sharing the same fluctuations
properties of their optical parameters (i.e., same matrix
$\mat{\cal C}$) but with distinct principal polarization
transformation vector $\g{p}_0$ can have a different depolarization
index, or Cloude entropy in the general case. This is due to the fact
that the depolarization index is calculated from the whole Mueller
matrix, and that the Cloude entropy depends on the four eigenvalues of
the CCM, i.e., both metrics simultaneously depend on the deterministic
polarization transformation and on the fluctuating parameters.

Contrarily, the new insight brought by the differential Jones and
Mueller calculus allows one to naturally define intrinsic
depolarization metrics, which only depend on the fluctuations of the
anisotropy parameters of the
sample%(or the depolarizing part of the dMm)
. One can first define the \emph{intrinsic differential depolarization
  metric} as $P_\delta=||\mat{ \Sigma}||_F$, where
$||\mat{X}||_F=\sqrt{\tr [\mat{X}^\dagger \mat{X}]}$ denotes the
Frobenius matrix norm \cite{ort15}. Such quantity can vary from $0$
(non-depolarizing) to (potentially) infinity and has proven to be
efficient in situations where standard approaches fail to correctly
describe the depolarizing nature of a sample \cite{ort15}.  In
addition, one can further gain a physical insight on the
depolarization properties of the sample by analyzing additional
quantities on the submatrix ${\Sigma}$. For instance, the determinant
of ${\Sigma}$ can be interpreted as a \emph{depolarization volume}
${\cal V}_{dep}=\det [\mat{\Sigma}]$. This quantity is equal to zero
as soon as one polarimetric direction has null fluctuations,
indicating perfect correlation between two polarization directions.
Another interesting approach is to analyze the Cloude entropy of the
submatrix $\mat{\Sigma}$ itself, which can be expressed as a
normalized version of the von Neumann entropy by
%\begin{equation}
$ {\cal S}({\Sigma})=-\tr \bigl[\frac{\mat{\Sigma}}{||\mat{\Sigma}||_1}\log_3
\frac{\mat{\Sigma}}{||\mat{\Sigma}||_1}\bigr]$,
%\end{equation}
where $||X||_1=\tr\sqrt{X^\dagger X}$ denotes the Schatten-von Neumann
1-norm (\emph{trace} norm) \cite{tom15}. The notation ${\cal S}$ is
used to avoid confusion with the Shannon entropy of the field
$H(\g{s})$ defined above. This quantity varies between 0 and 1 and
adds interesting information on the distribution of the eigenvalues of
$\mat{\Sigma}$, which informs about depolarization anisotropy.  It is
interesting to notice that the quantities $P_\delta$, ${\cal V}_{dep}$
and ${\cal S}({\Sigma})$ are defined irrespective of the propagation
distance, and are invariant by deterministic unitary transformations,
thus justifying their qualification as intrinsic metrics. This has the
strong physical meaning that the sample must keep the same
depolarization properties whatever be its deterministic anisotropic
properties. Moreover, combining these three depolarization parameters
provides direct information on the number and degeneracy of non-null
canonical parameters, as evidenced in Table~\ref{tab_dep_param}.  Such
a procedure, which does not require reducing $\mat{m}_d$ to its
canonical form, also allows one to identify
type-$(\textsc{\romannumeral 1})$ canonical family when all three
canonical parameters are non-null.

\begin{table}[t]
\begin{tabular}{c|c|ccc}
\multicolumn{2}{c|}{Parameters $d_{i,\ i\in \{1,3\}}$} & $P_\delta^2$ & ${\cal V}_{dep}$ &$ {\cal S}({\cal C})$\\
\hline
(\textsc{\romannumeral 1})&$d_i=d_j=d_k$ & $3d_i$ &$d^3_i$&1\\
&$d_i=d_j\neq d_k$ &  $2d_i +d_k $&$d^4_id_k$&$\frac{\ln2}{\ln3}<{\cal S}<1$\\
&$d_i\neq d_j\neq d_k$ & $d_i + d_j+d_k $ &$d_id_jd_k$&$0<{\cal S}<1$\\
\hline
(\textsc{\romannumeral 1}/\textsc{\romannumeral 2})&$d_i = d_j \quad   0 $ &$2d_i $ &$0$&$\frac{\ln2}{\ln3}$\\
&$d_i \neq  d_j \quad   0 $ & $d_i + d_j$& $0$&$0<{\cal S}<\frac{\ln2}{\ln3}$\\
&$d_i \quad 0 \quad  0 $ &$d^2_i$ &$0$&$0$\\
%&$0 \quad 0 \quad  0 $ & $0$&$0$&$-$\\
\end{tabular}
\caption{Intrinsic differential depolarization metric, volume and Cloude entropy for the two canonical dMm's of Eq.~(\ref{canon}). \label{tab_dep_param}}
\end{table} 

These considerations provide a fundamental insight on the origin of
depolarization as a randomization of light polarization due to the
second-order statistical fluctuations of the anisotropy parameters of
the medium, giving access to meaningful intrinsic depolarization
metrics. Lastly, they allow us to demonstrate an irreversibility
property of depolarizing light-matter interactions:
\begin{prop}\label{th1}
  For any admissible fully or partially polarized input Stokes vector
  $\g{s}_{in}$, a physically realizable depolarizing non-singular
  and unit determinant Mueller matrix $\mat{\widetilde{M}}$ verifies 
%\begin{equation}\label{cond}
$\mink{\g{s}_{out}}^2 = \mink{\mat{\widetilde{M}}\,\g{s}_{in}}^2 \geq \mink{\g{s}_{in}}^2$.
%\end{equation}
\end{prop}
The demonstration of this property in the general case of standard
Mueller matrices has never been reported to our best knowledge, and is
provided in a more general form as Supplemental Material
\cite{SI}. For the sake of conciseness, we provide the demonstration
of the equivalent property for a depolarizing dMm
$\mat{m}=\mat{m}_{nd}+\mat{m}'_d$ with null trace ($\kappa_i=0$),
i.e., for any physical Stokes vector,
\begin{equation}\label{irre}
\frac{d \mink{\g{s}}^2}{dz} = \g{s}^T \bigl[ {\mat{m}'}_d^{T}\, \mat{G} + \mat{G}\, \mat{m}'_d \bigr]\g{s}\geq 0,
\end{equation}
with equality if and only if the dMm is non depolarizing
($\mat{m}'_d=0$). The above expression is easily obtained by first
order Taylor expansion of
$\mink{\g{s}(z+\Delta z)}^2=\mink{(\Id+ \mat{m} \Delta z )\g{s}(z)}^2$
and by the G-antisymmetry of $\mat{m}_{nd}$ when $\kappa_i=0$. The
positivity can then be easily shown on the two canonical forms of
$\mat{m}'_d$ recalled above. One indeed has
$d \mink{\g{s}}^2/dz=\Delta z [2 s_0^2(d_1+d_2+d_3)-s_1^2
(d_1-d_2-d_3)-s_2^2(d_2-d_3-d_1)-s_3^2(d_3-d_1-d_2)]$ for
type-$ (\textsc{\romannumeral 1})$ depolarizing dMm, whereas, for
type-$ (\textsc{\romannumeral 2})$,
$d \mink{\g{s}}^2/dz=\Delta z[
2d_2(s_0^2+s_1^2)+d_1(s_0^2-s_1^2+s_2^2+s_3^2)]$. These two quantities
are obviously non negative under the physicality conditions recalled
above (i.e., $\forall i\in\{1,3\},\ d_i \geq 0$) and for admissible
Stokes vectors (i.e., $\mink{\g{s}}\geq0$).

\begin{figure}[t]
\begin{center}
\includegraphics[width=7.7cm]{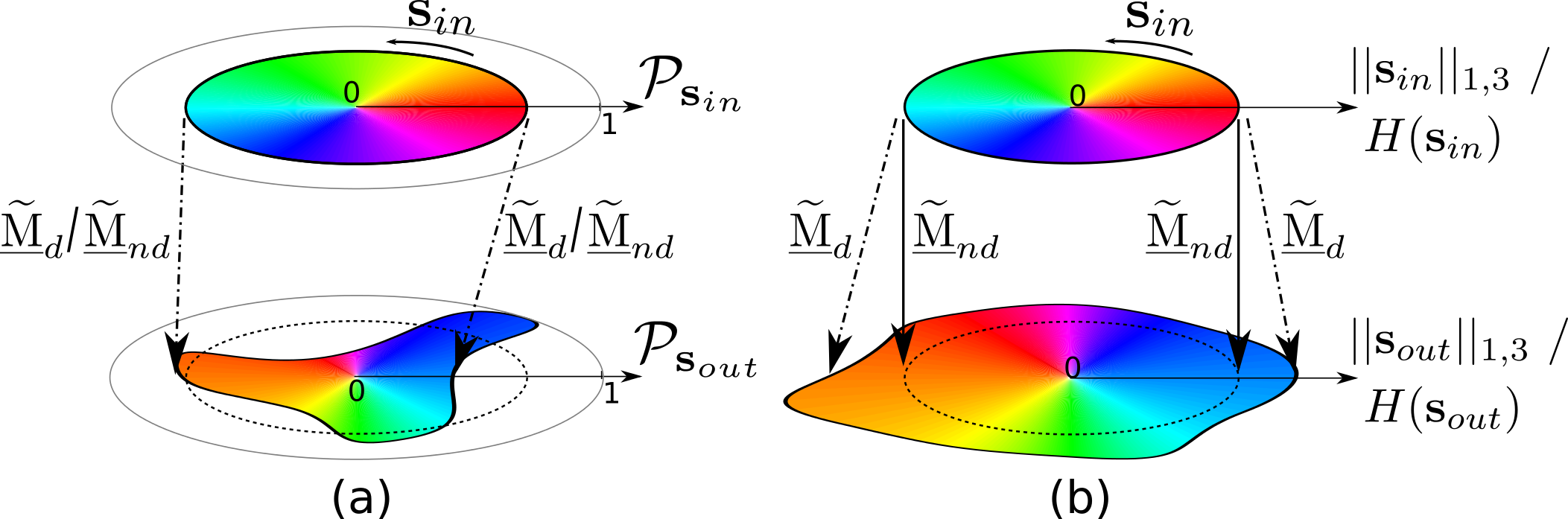}
\end{center}
\caption{Effect of a non-depolarizing
  ($\widetilde{M}_{nd}$) or a depolarizing ($\widetilde{M}_{d}$)
  transformation on (a) the standard degree of polarization  ${\cal P}$, and (b) the Minkowski metric $||\g{s}||_{1,3}$ or
  Shannon entropy $H(\g{s})$, for all possible input Stokes vectors
  $\g{s}_{in}$, schematically represented around a chromatic
  disk. % Contrarily to the DOP, $||\g{s}||_{1,3}$ or $H(\g{s})$ are
  % preserved for non-depolarizing (deterministic and reversible)
  % transformations, and must necessarily grow for depolarizing (random
  % and irreversible) transformations.
  \label{conserv}}
\end{figure}

Property \ref{th1} has a strong physical meaning since it reveals the
irreversible effect of a depolarizing transformation on the
propagating field, which clearly appears through the necessary
increase of the Minkowski metric of its Stokes vector. Interestingly,
this irreversibility property has an
\emph{informational/thermodynamical} counterpart. Indeed, from
Eq.~(\ref{irre}), the Shannon entropy of the bidimensional electrical
field vector can be shown to obey an irreversible evolution with
depolarizing transformations, as
%\begin{equation}
$\frac{d H(\g{s})}{dz}=\frac{2}{\mink{\g{s}}}\frac{d \mink{\g{s}}}{dz}
  \geq 0$.
%\end{equation}
  Such an irreversible behavior of the Minkowski metric
  $||\g{s}||_{1,3}$, or alternatively the Shannon entropy $H(\g{s})$,
  confirms that these quantities are best adapted to describe the
  polarimetric \emph{randomization} (depolarization) of a propagating
  beam. Contrarily to the field intensity or the standard degree of
  polarization $\degp$, these quantities are preserved through
  non-singular deterministic (and reversible) transformations, and
  must necessarily grow with irreversible depolarizing
  transformations. This is schematically illustrated in
  Fig.~\ref{conserv}.

% \begin{figure}[ht]
% \begin{center}
% \includegraphics[width=7cm]{sorting_matrices}
% \end{center}
% \caption{The physically realizable non-singular Mueller matrices with
%   unit determinant represent a subset of the real $4\times 4$
%   matrices. Each of these matrices is associated to a dMm that can be
%   splitted into a non-depolarizing dMm
%   $\mat{m}_{nd} \in \mathfrak{so}^+(3,1)$ and a physically realizable
%   depolarizing dMm $\mat{m}'_d$. The physically realizable Mueller
%   matrices and dMms must necessarily increase the Minkoswki metric of
%   any input Stokes vector.}
% \end{figure}

  As a conclusion, the differential Jones formalism has allowed us to
  provide a clear intrinsic physical picture of the origins of light
  depolarization in a medium, as well as physically meaningful
  implication regarding an irreversibility property of the beam
  entropy under depolarizing transformations. % An interesting

\end{document}